# Counting individual $^{41}$Ca atoms with a Magneto-Optical Trap

I.D. Moore,* K. Bailey, J. Greene, Z.-T. Lu,* P. Müller, T. P. O'Connor

*Physics Division, Argonne National Laboratory, Argonne, Illinois 60439*

Ch. Geppert, K.D.A. Wendt*

*Institut für Physik, Johannes Gutenberg-Universität Mainz, D-55099 Mainz, Germany*

L. Young

*Chemistry Division, Argonne National Laboratory, Argonne, Illinois 60439*

Atom Trap Trace Analysis (ATTA), a novel method based upon laser trapping and cooling, is used to count individual atoms of $^{41}$Ca present in biomedical samples with isotopic abundance levels between $10^{-8}$ and $10^{-10}$. ATTA is calibrated against Resonance Ionization Mass Spectrometry, demonstrating a good agreement between the two methods. The present ATTA system has a counting efficiency of $2\times10^{-7}$. Within one hour of observation time, its 3-$\sigma$ detection limit on the isotopic abundance of $^{41}$Ca reaches $4.5\times10^{-10}$.

Calcium is one of the most abundant elements on earth, and is of vital importance for the metabolism of biological organisms. Consequently, the analysis of its long-lived radioactive isotope, $^{41}$Ca ($t_{1/2}$ = 1.03×10$^5$ year), has important applications in both earth and life sciences. On earth, $^{41}$Ca is produced predominantly as a cosmogenic isotope via the $^{40}$Ca(n,γ)$^{41}$Ca reaction [1], resulting in a natural isotopic abundance in the range of 10$^{-15}$ – 10$^{-14}$ on the Earth's surface. Hence, $^{41}$Ca is a candidate for dating bones ranging from fifty thousand to one million years of age [2,3]. This period is particularly important in the archeological study of early human development, and is beyond the reach of $^{14}$C-dating [4]. $^{41}$Ca can also be used in geology to determine the exposure ages of rocks or in cosmochemistry for investigations on terrestrial ages and shielding of meteorites [5]. Moreover, artificial $^{41}$Ca can be used in studies of calcium metabolism in living systems. One interesting example is to use $^{41}$Ca tracer to directly monitor the bone-loss and retention rates of human subjects in both research and diagnosis of osteoporosis [6,7]. In biomedical applications the long half-life of $^{41}$Ca translates into low specific radioactivity, and the isotope tracer can be safely introduced into subjects at an initial isotopic abundance as high as 10$^{-8}$.

Approaches to analyze $^{41}$Ca at the natural level have been demonstrated using Accelerator Mass Spectrometry (AMS) at several high-end (energy ~ 10 MeV) facilities [2,5]. However, the results obtained so far are not definitive because the natural level is very close to the detection limit (~ 10$^{-15}$–10$^{-13}$) of AMS. For biomedical applications, where the isotopic abundance can be raised to the level of 10$^{-13}$–10$^{-8}$, $^{41}$Ca analyses have been successfully conducted at several AMS facilities [8]. More recently, Resonance Ionization Mass Spectrometry (RIMS) [9,10], a method combining the selective power of both laser spectroscopy and mass spectrometry, has been applied to analyze $^{41}$Ca in environmental and biomedical samples [11]. A detection limit of RIMS as low as ~ 10$^{-13}$ has been demonstrated with metallic Ca samples. With biomedical samples, the 3-σ limit is presently at $4.3\times10^{-11}$ due to the interference arising from a high concentration of $^{41}$K, which cannot be fully removed during chemical sample preparation [12,13]. Compared with AMS, RIMS uses a much smaller apparatus and costs significantly less, both advantages are important for practical biomedical applications.

We report in this paper the first detection and analysis of $^{41}$Ca using Atom Trap Trace Analysis (ATTA), a method based upon the techniques of manipulating and detecting neutral atoms with resonant laser light. In ATTA, individual atoms of the desired trace-isotope are selectively cooled in a Zeeman slower and trapped in a magneto-optical trap (MOT) by resonant laser beams, and are detected by observing the fluorescence of the trapped atoms. The principle of ATTA was demonstrated earlier by Chen *et al*. with the successful analysis of the $^{81}$Kr/Kr ratio (~ $10^{-13}$) in atmospheric samples [14]. ATTA shares with RIMS the advantages of lower cost and smaller apparatus, and has the potential of achieving the selectivity required to analyze $^{41}$Ca at the natural level.

In this work, laser cooling and trapping of neutral calcium atoms are performed by resonantly exciting the $4s^2\ ^1S_0 \rightarrow 4s4p\ ^1P_1$ transition (natural linewidth = 34.6 MHz). The required laser beams, with a total power of approximately 80 mW at 422.7 nm, are produced by a frequency-doubled cw Ti:Sapphire ring laser whose frequency is referenced to a stabilized Fabry-Perot cavity. The schematic of the atomic beam system is shown in Fig. 1. In an analysis, a sample in the form of either metallic calcium or an inorganic compound such as calcium nitrate, $Ca(NO_3)_2$, is loaded into an oven with a 2 cm long, 2 mm diameter nozzle. The oven can be heated up to 1000°C to produce a collimated calcium atomic beam. To reduce the abundant $^{40}$Ca beam flux the isotope of interest is selectively deflected by approximately 5°, and transversely cooled in that direction with a pair of laser beams before entering a Zeeman slower and finally being captured by a MOT. We note that each of these laser-manipulation steps is isotopically selective.

By tuning the laser frequency within a few natural linewidths on the low-frequency side of the resonance of a particular isotope, only atoms of this isotope are trapped. Atoms of other isotopes are either deflected before reaching the trap or pass through without being captured. The number of trapped atoms is determined by measuring their fluorescence. Atoms remain trapped for an average of 18 ms, which is limited by a weak decay channel from the excited 4s4p $^1P_1$ level, through intermediate levels, to the metastable 4s4p $^3P_2$ level. All of the six stable calcium isotopes have been trapped and observed in this system [15]. When a metallic calcium sample is heated to 570 °C, the trap can capture the abundant $^{40}$Ca (97%) at a rate of $5\times10^8$ atoms/s with a capture efficiency of $3\times10^{-5}$. When a calcium nitrate sample is used, the oven temperature must be raised to 750 °C in order to reduce the molecules to atoms. As a result of the higher temperature the capture rate is lowered to $1\times10^8$ $^{40}$Ca atoms/s and the efficiency to $2\times10^{-7}$.

In order to observe the rare isotope $^{41}$Ca, the system must be sensitive enough to resolve and count single atoms at a sufficient loading rate such that an abundance measurement may be made within a practical time period. The frequency difference between the abundant $^{40}$Ca isotope and $^{41}$Ca [16], amounts to only 155 MHz, corresponding to 4.4 times the natural linewidth. Despite having selectively deflected the atomic beam upon exiting the oven, the dominant source of background arises from the scattering of photons off thermal $^{40}$Ca atoms passing through the detection region. Figure 2 shows a typical example of the raw data that indicates the fluorescence of individual $^{41}$Ca atoms in the trap. Over a typical trap lifetime of 18 ms, an atom scatters photons at the rate of $3\times10^6$ s$^{-1}$, 2.5% of which are imaged and counted by a photo-multiplier detector with an integration time of 8 ms. The signal size of an atom is dependent on the duration of time the atom spends in the trap. The average background photon count rate of 209 photons/8 ms is due to light scattered off the $^{40}$Ca atomic beam. A threshold condition, indicated by the dashed line, of $5\sigma$ (= 80 photons/8 ms) above the mean (= 209 photons/8 ms) of the background is required for an event to be counted as a single atom. The choice of this threshold setting is determined by the statistical distribution of background and single atom data. The signal-to-noise ratio of the largest peak in figure 2 is 17. In order to ensure that we have indeed detected $^{41}$Ca and not one of the abundant isotopes, we mapped out atom counts as a function of the laser frequency (Fig. 3). For comparison the nearest stable isotopes $^{40}$Ca (96.9%), $^{42}$Ca (0.65%) and $^{43}$Ca (0.14%) were also trapped successively and are shown in the upper trace. The peak of $^{41}$Ca atom counts occur at 166 MHz above the trap fluorescence of $^{40}$Ca, which agrees with a previous spectroscopic measurement [16]. Moreover, the absence of counts on both sides of the $^{41}$Ca peak during the measurement duration of six hours demonstrates that interference from the neighboring abundant isotopes is suppressed to below an isotopic abundance level of $7\times10^{-10}$.

We have analyzed three biomedical samples and compared the ATTA results with those of RIMS, which in turn had previously been calibrated with AMS measurements. The samples were taken and provided by partners of the Osteodiet research project of the European Community [17]. In this program, subjects were given a 100 nCi dose of $^{41}$Ca. Urine samples were taken starting six days after ingestion up to later periods of 100 days and more. At the ETH Zürich, Switzerland, these raw samples were chemically prepared into the form of a 3M calcium nitrate solution, of which a 10 μl drop contains approximately $1\times10^{18}$ calcium atoms. The $^{41}$Ca/Ca ratios were measured using RIMS at the University of Mainz and using ATTA at Argonne National Laboratory. For each ATTA measurement 40 μl of the solution is absorbed and dried on a titanium sponge, while the RIMS measurement uses 10 μl of solution on a titanium foil. The titanium acts as an efficient reducing agent for the nitrate.

In order to measure the isotopic abundance of $^{41}$Ca, the ATTA system is continuously switched between 2 minutes of trapping $^{41}$Ca and 10 seconds of trapping $^{42}$Ca for normalization. While single atom detection is performed to count $^{41}$Ca, the $^{42}$Ca trap typically contains $10^3$ atoms, whose fluorescence has to be reduced by a filter before being measured with the same photon counter. At the end of a measurement, which typically lasts for three hours, the counts for each isotope are summed up and a ratio of counts between the two isotopes is derived. Note that for precise isotope ratio measurements, the system has to first be calibrated with samples of known ratios, which in this case are the RIMS values. Three samples (Z1, Z2 and Z3) were measured using both ATTA and RIMS (Fig. 4). A fourth sample of non-enriched calcium nitrate solution was used in a null measurement using ATTA. The measured isotope ratios are converted to $^{41}Ca/Ca_{total}$ by using the known isotope ratios of stable reference Ca isotopes. The errors on the ATTA measurements are dominated by a ~10% statistical error on the $^{41}Ca$ counts. The linear correlation between the ATTA and RIMS values shown in Fig. 4 not only serves as a calibration of ATTA, but also demonstrates the validity of ATTA as a quantitative analysis tool. From the absence of counts during the null sample measurement, we conclude that within one hour of observation the 3-$\sigma$ detection limit on $^{41}Ca/Ca$ reaches $4.5 \times 10^{-10}$.

In conclusion, we have demonstrated a new method of analyzing $^{41}Ca/Ca$ ratios in biomedical samples. Significant improvements to the ATTA system are required in order to determine much lower isotope abundances, e.g. for dating bones at the natural abundance level. The use of metallic calcium will allow the oven to hold a larger sample, and enable the system to achieve a higher loading rate as well as a higher trap efficiency. Additional increase can be realized by implementing a full two-dimensional transverse cooling. In order to reduce the background due to atomic beam fluorescence the trap lifetime would need to be increased to enable temporal separation of trap loading and detection, in a similar way as has been used for $^{81}Kr$ analyses using ATTA [14]. Preliminary tests indicate that this increase in trap lifetime can be achieved by repumping the intermediate 4s3d $^1D_2$ level with a diode laser at 672 nm wavelength.

We thank Y.M. Li for contribution in the early stages. We also thank I. Ahmad and K. Orlandini for help in sample preparation. This work is supported by the U.S. Department of Energy, Nuclear Physics Division, and L.Y. is supported by the Office of Basic Energy Sciences, Division of Chemical Sciences, under contract W-31-109-ENG-38. This work was carried out with financial support from the Eurpean Commision Quality of Life Fifth Framework Programme QLK1-CT-1999-00752.

* To whom correspondence should be addressed. Email: lu@anl.gov; moore@mep.phy.anl.gov; klaus.wendt@uni-mainz.de.

Figure 1. Schematic of the calcium ATTA system.

Figure 2. Fluorescence of individual trapped $^{41}$Ca atoms. The threshold for accepting a single atom count is indicated by the dashed line.

Figure 3. Trap fluorescence versus laser frequency. The spectrum on $^{41}$Ca is accumulated over a six-hour period using a sample with a $^{41}$Ca/Ca ratio of $1\times10^{-8}$.

Figure 4. Comparison between ATTA and RIMS results on biomedical samples. A best-fit line to the data yields a reduced chi-squared value of 1.0.

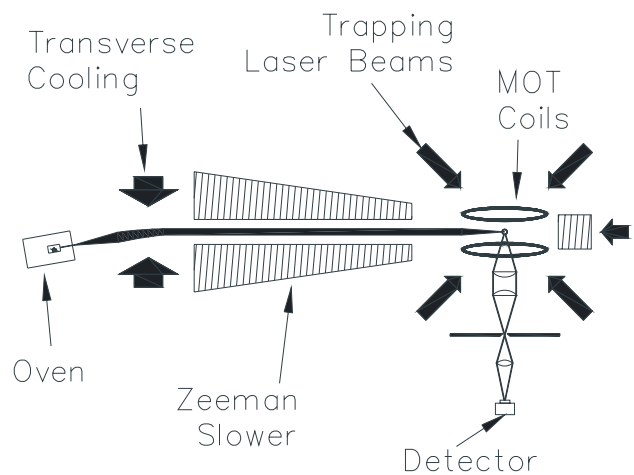

Figure 1

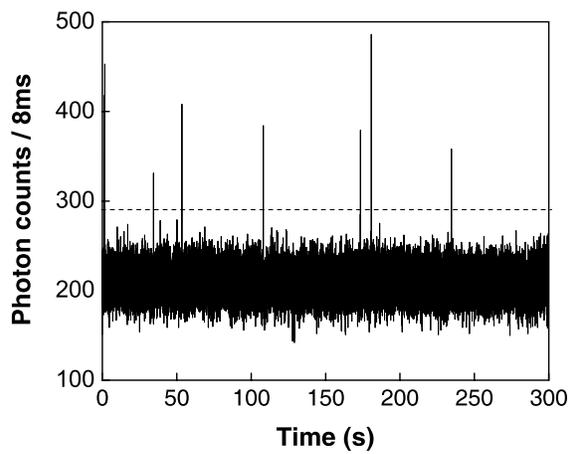

Figure 2

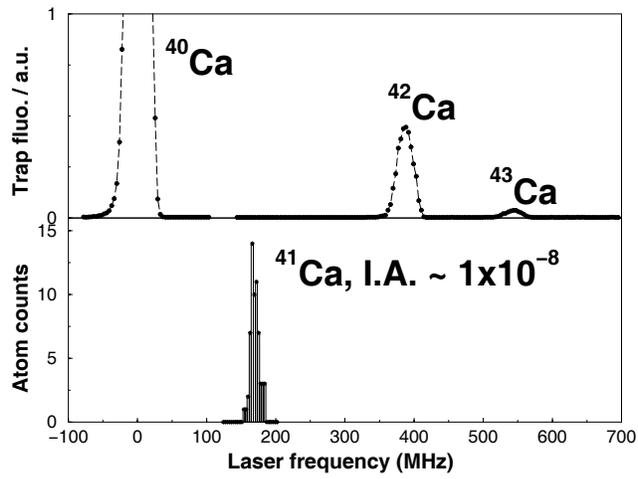

Figure 3

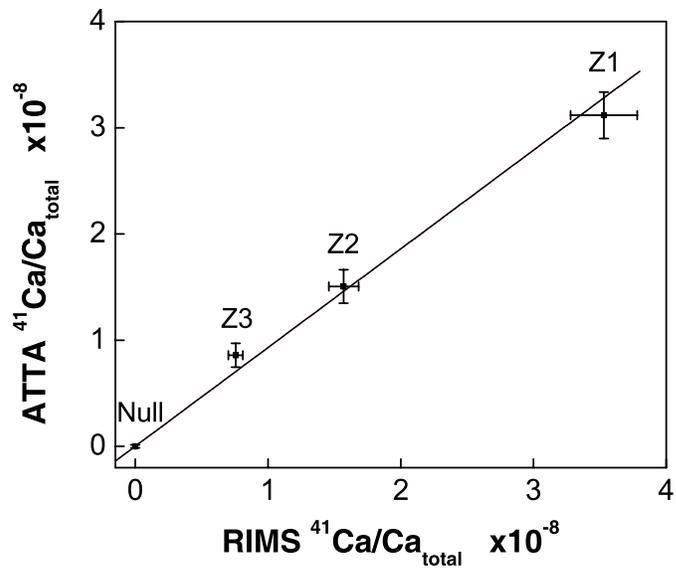

Figure 4